\documentclass[prl,superscriptaddress,twocolumn,showpacs,preprintnumbers,amsmath,amssymb]{revtex4}

\usepackage{graphicx}
\usepackage{dcolumn}
\usepackage{bm}

\begin{document}

\preprint{APS/123-QED}

\title{An Isotopic Fingerprint of Electron-Phonon Coupling in High-$T_{\mathrm c}$ Cuprates}

\author{H. Iwasawa}
\thanks{Present address: Hiroshima Synchrotron Radiation Center, Hiroshima University, Higashi-Hiroshima 739-8526, Japan}
\affiliation{Department of Applied Physics, Tokyo University of Science, Shinjuku-ku, Tokyo 162-8601, Japan}
\affiliation{National Institute of Advanced Industrial Science and Technology, Tsukuba, Ibaraki 305-8568, Japan}

\author{J. F. Douglas}
\affiliation{Department of Physics, University of Colorado, Boulder, CO 80309-0390, USA}

\author{K. Sato}
\affiliation{National Institute of Advanced Industrial Science and Technology, Tsukuba, Ibaraki 305-8568, Japan}
\affiliation{Faculty of Science, Ibaraki University, Mito, Ibaraki 310-8512, Japan.}

\author{T. Masui}
\affiliation{Department of Physics, Osaka University, 1-1 Machikaneyama, Toyonaka, Osaka 560-0043, Japan}

\author{Y. Yoshida}
\affiliation{National Institute of Advanced Industrial Science and Technology, Tsukuba, Ibaraki 305-8568, Japan}

\author{Z. Sun}
\affiliation{Department of Physics, University of Colorado, Boulder, CO 80309-0390, USA}

\author{H. Eisaki}
\author{H. Bando}
\affiliation{National Institute of Advanced Industrial Science and Technology, Tsukuba, Ibaraki 305-8568, Japan}

\author{A. Ino}
\affiliation{Graduate School of Science, Hiroshima University, Higashi-Hiroshima 739-8526, Japan}

\author{M. Arita}
\author{K. Shimada}
\author{H. Namatame}
\affiliation{Hiroshima Synchrotron Radiation Center, Hiroshima University, Higashi-Hiroshima 739-8526, Japan}

\author{M. Taniguchi}
\affiliation{Graduate School of Science, Hiroshima University, Higashi-Hiroshima 739-8526, Japan}
\affiliation{Hiroshima Synchrotron Radiation Center, Hiroshima University, Higashi-Hiroshima 739-8526, Japan}

\author{S. Tajima}
\affiliation{Department of Physics, Osaka University, 1-1 Machikaneyama, Toyonaka, Osaka 560-0043, Japan}

\author{S. Uchida}
\affiliation{Department of Physics, University of Tokyo, Tokyo 113-8656, Japan}

\author{T. Saitoh}
\affiliation{Department of Applied Physics, Tokyo University of Science, Shinjuku-ku, Tokyo 162-8601, Japan}

\author{D. S. Dessau}
\affiliation{Department of Physics, University of Colorado, Boulder, CO 80309-0390, USA}

\author{Y. Aiura}
\email[To whom all correspondence should be addressed.\\ E-mail: ]{y.aiura@aist.go.jp}
\affiliation{National Institute of Advanced Industrial Science and Technology, Tsukuba, Ibaraki 305-8568, Japan}
\affiliation{Hiroshima Synchrotron Radiation Center, Hiroshima University, Higashi-Hiroshima 739-8526, Japan}

\date{\today}

\begin{abstract}

Angle-resolved photoemission spectroscopy with low-energy tunable photons along the nodal direction of oxygen isotope substituted Bi$_2$Sr$_2$CaCu$_2$O$_{8+\delta}$ reveals a distinct oxygen isotope shift near the electron-boson coupling $``$kink" in the electronic dispersion. 
The magnitude (a few meV) and direction of the kink shift are as expected due to the measured isotopic shift of phonon frequency, which are also in agreement with theoretical expectations.
This demonstrates the participation of the phonons as dominant players, as well as pinpointing the most relevant of the phonon branches.

\end{abstract}


\maketitle


Effects of electron-boson interactions on electronic self-energies show up as sudden changes in the electron dispersion, or $``$kinks", via angle-resolved photoemission spectroscopy (ARPES) \cite{Damascelli03, Campuzano04}. 
While the kinks are now indisputable, their origin as arising from electronic coupling to phonons \cite{Lanzara01, Shen02, Cuk04}, magnetic excitations \cite{Scalapino99, Carbotte99, He02, Terashima06}, or both\cite{Gromko03}, remains unclear. 
The best way to identify the origin of an interaction is to slightly modify the interaction in a controlled manner. 
For phonons, this can be done via an isotopic exchange, which varies particle masses and hence the vibrational energies. 
Here we substitute the oxygen isotopes ($^{16}$O $\rightarrow$ $^{18}$O) leading to a softening of phonon energies of a few meV [=$\Omega\times(1$$-$$\sqrt{\smash[b]{16/18}})$] where $\Omega$ are the bare phonon frequencies of 40-70 meV \cite{Reznic95, McQueeney01}.
Recent scanning tunnelling microscopy (STM) experiments showed an isotopic shift of a second derivative feature of 3.7 meV \cite{Lee06}, though whether the observed shift is evidence for electron-phonon coupling has been disputed by a number of groups in terms of the inelastic tunnelling barrier \cite{Pilgram06, Scalapino06, Hwang07}. 
Even earlier than that, Gweon {\em et al.} used ARPES to study isotope effects on the electronic structure \cite{Gweon04}.  
They found that the primary effect was at high energy and was unusually strong (up to 30 meV), though these results were not reproduced by more modern experiments \cite{Douglas07, Iwasawa07}.
Therefore, a direct and clear $``$fingerprint" of strong electron-phonon interactions in the high-$T_{\mathrm c}$ superconductors has been missing up to now.

The momentum-selectivity of ARPES brings an additional tool to bear on the study of the isotope effect which is not available from tunneling. 
This information is very helpful in separating out the energy and momenta of the bosonic modes which couple most strongly to the electronic degrees of freedom.
Here we focus on the $``$nodal" spectra, i.e. those along the (0, 0)-($\pi$, $\pi$) direction in the Brillouin zone where both the superconducting gap and pseudogap are minimal or zero \cite{Damascelli03, Campuzano04}. 
This is the portion of the Brillouin zone where the kinks or self energy effects have been most strongly studied both experimentally \cite{Lanzara01, Valla99, Zhou03} and theoretically \cite{Zhang07, Kulic07}. 
As described very recently by Giustino {\em et al.} \cite{Giustino08} and Heid {\em et al.} \cite{Heid08}, there is still great controversy about the nature of the coupling in this direction; these two theoretical arguments imply that the phonon coupling could only be a minor player in the overall electron-boson coupling of these states because the calculations indicate a significantly weaker kink than what is found in experiment. 

Analysis and interpretation of the nodal data is also more straightforward than the data throughout the Brillouin zone, as the energies of any bosonic modes in the spectra are expected to be shifted by the energies of the gaps. 
Assuming any isotope shifts go roughly as [$\Omega\times(1$$-$$\sqrt{\smash[b]{16/18}})$], these are at most a few meV, requiring an unusually high precision ARPES experiment. 
To verify such a tiny effect in the most convincing way, we utilized bulk-sensitive low photon energy ARPES (LE-ARPES), which has greatly improved spectral resolution compared to $``$conventional" ARPES \cite{Koralek06}. 
In this Letter, we report an isotopic fingerprint of electron-phonon coupling as a clear isotope shift of the kink energy taking advantages of LE-ARPES.
Present results provide straightforwardly that the origin of the nodal kink is due to the electron-phonon interactions. 


\begin{figure*}
	\begin{center}
	\includegraphics[width=120 mm, keepaspectratio]{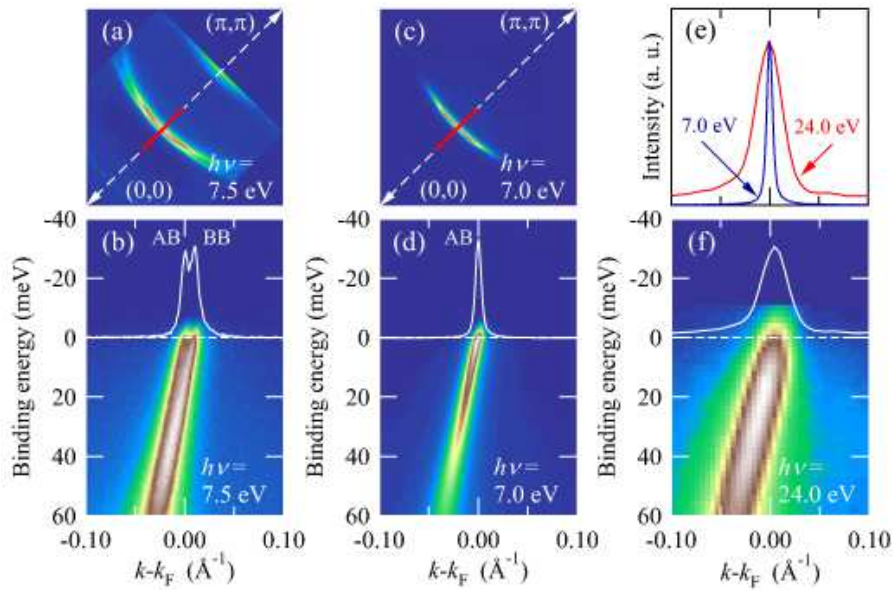}
	\end{center}
	\caption{(color online).
(a) and (c) Fermi surface maps taken with 7.5 eV and 7.0 eV photons, respectively. 
(b), (d), (f) Energy-momentum distribution maps taken with 7.5 eV, 7.0 eV and 24.0 eV along the nodal direction, indicated by the red line in (a) and (c). 
MDCs at the Fermi level ($E_{\rm F}$) are also shown by the white line, where the BB and AB denote bonding band and antibonding band, respectively. 
The Fermi momentum is set to the momentum of the AB crossing $E_{\rm F}$. 
(e) Comparison of MDCs at $E_{\rm F}$ between 7.0 eV (blue) and 24.0 eV (red).
Note that ARPES data using a conventional photon-energy of 24.0 eV were taken at BL5-4 of the Stanford Synchrotron Radiation Laboratory (SSRL).} 
\label{FIG1}
\end{figure*}

High-quality optimally doped Bi$_2$Sr$_2$CaCu$_2$O$_{8+\delta}$ (Bi2212) single crystals were prepared by the travelling-solvent floating-zone technique \cite{Eisaki04}. 
Oxygen isotope substitution was performed by annealing procedures \cite{Anneal}, yielding a slight decrease of $T_{\mathrm c}$ from 92.1 K to 91.1 K.
A softening of the oxygen vibration modes and the high oxygen isotope substitution rate (more than 80 $\%$) was confirmed by Raman spectroscopy \cite{Raman}.
Further, LE-ARPES gives us unparalleled precision in determining the Fermi surface areas directly from the Fermi surface maps, and we have confirmed that these are the optimal and the same for the two isotopes to the level below 2 $\%$ (a doping level uncertainty $\mathit{\Delta}x$=.003).

Present data were collected at BL-9A of Hiroshima Synchrotron Radiation Center using a Scienta R4000 electron analyzer. 
We used our newly developed high-precision 6-axis sample manipulator to remove the extrinsic effect due to a sample misalignment. 
The clean and flat surface of the samples was obtained by cleaving {\em in situ} in ultrahigh vacuum better than 4 $\times$ 10$^{-11}$ Torr below 10 K. 
The total instrumental energy and angular (momentum) resolutions were better than 5 meV and 0.4$^\circ$ (0.005 $\textrm \AA^{-1}$) at 7.0 eV photons, respectively. 

First, we optimized the excitation energy to avoid the complications in extracting accurate band dispersions due to the very small but finite (0.01 $\textrm {\AA}^{-1}$) bilayer-splitting which is now known to exist even along the nodal line [Figs.~\ref{FIG1}  (a) and (b)] \cite{Yamasaki07}. 
By tuning the photon energy to 7.0 eV, we can see the complete isolated and individual antibonding band dispersion [Figs.~\ref{FIG1}  (c) and (d)], giving the extremely fine quasiparticle dispersion with the momentum full width ($\mathit{\Delta}k$) below 0.005 $\textrm {\AA}^{-1}$ [blue line in Fig.~\ref{FIG1} (e)]. 
This is in sharp contrast to the cases of the dispersion with bilayer-splitting [Fig.~\ref{FIG1}  (b)], and also to the conventional ARPES spectra broadened by a limit of resolutions as well as the unresolved bilayer-splitting [Fig.~\ref{FIG1} (f)]. 
These advantages of LE-ARPES, including the band selectivity, are mandatory to verify the subtle change in the ARPES spectra with isotope substitution.

\begin{figure}
	\begin{center}
	\includegraphics[width=70 mm, keepaspectratio]{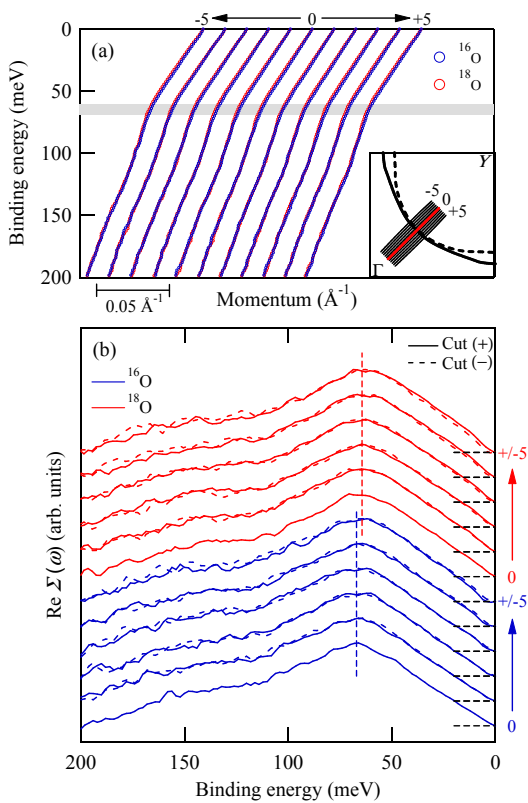}
	\end{center}
	\caption{(color online). (a) Energy-momentum dispersions near the nodal region both for $^{16}$O (blue) and $^{18}$O (red) from optimally doped Bi2212. 
Those measured cuts are labeled -5 to +5, displayed in the inset. 
Gray shaded are roughly indicate the kink in the all dispersions. 
(b) Real-parts of the self-energy both for $^{16}$O (blue) and $^{18}$O (red), showing a isotope shift of $\sim$ 70 meV peak. 
Blue and red dashed line indicates the kink energy, averaged from cut -5 to cut +5, for $^{16}$O and $^{18}$O, respectively. 
Positive and negative cuts are represented by solid and dashed lines, respectively.}	
\label{FIG2}
\end{figure}

Figure~\ref{FIG2} (a) compares the $^{16}$O (blue) and $^{18}$O (red) energy-momentum dispersions near the nodal region taken with 7.0 eV photons. 
We found a very small isotope effect around the kink in the dispersions, independent of the cut position. 
We emphasize here that each of the five positive dispersions is exactly identical with each of the five negative dispersions for both $^{16}$O and $^{18}$O [as seen in Fig.~\ref{FIG2} (b)], which rules out the possibility of an extrinsic effect due to a sample misalignment. 
An earlier attempt at studying isotopic shifts by conventional ARPES showed many unusual effects of the oxygen substitution, characterized by extremely large (30 meV scale) shifts of the high binding energy portions of the spectra \cite{Gweon04}. 
These unusual isotope effects were not verified in recent ARPES experiments \cite{Douglas07, Iwasawa07} nor by the ultra-high resolution experiments shown here.
It should be noted that the present results are consistent with the previous our ARPES \cite{Douglas07, Iwasawa07}, though we can not see a convincing evidence of such a tiny isotope effect in there.
This should be attributed to that the experimental resolutions and/or accuracy were still not enough to observe a few meV order effect.

To visualize the subtle isotope effect more clearly, we deduced the real part of the self-energy Re$\mathit\Sigma$($\mathit\omega$).
We extracted Re$\mathit\Sigma$($\mathit\omega$) by subtracting a bare band dispersion from the experimental one using three different forms of bare band in this analysis: linear (presented in the figure) \cite{Johnson01}, second order polynomial \cite{Meevasana06}, and third order polynomial. 
An important thing is that the present result is robust in regard to this choice. 
As seen in Fig.~\ref{FIG2} (b), clearly observed is a few meV isotopic shift of the $\sim$ 70 meV peak of Re$\mathit\Sigma$($\mathit\omega$).
Here we estimated the kink energy as the energy giving the peak maximum of Re$\mathit\Sigma$($\mathit\omega$), which was obtained by fitting the top portion of Re$\mathit\Sigma$($\mathit\omega$) with a Gaussian. 
To quantify the energy-scale of the kink as well as the isotopic kink shift, we have studied multiple samples systematically. 

\begin{figure*}
	\begin{center}
	\includegraphics[width=120 mm, keepaspectratio]{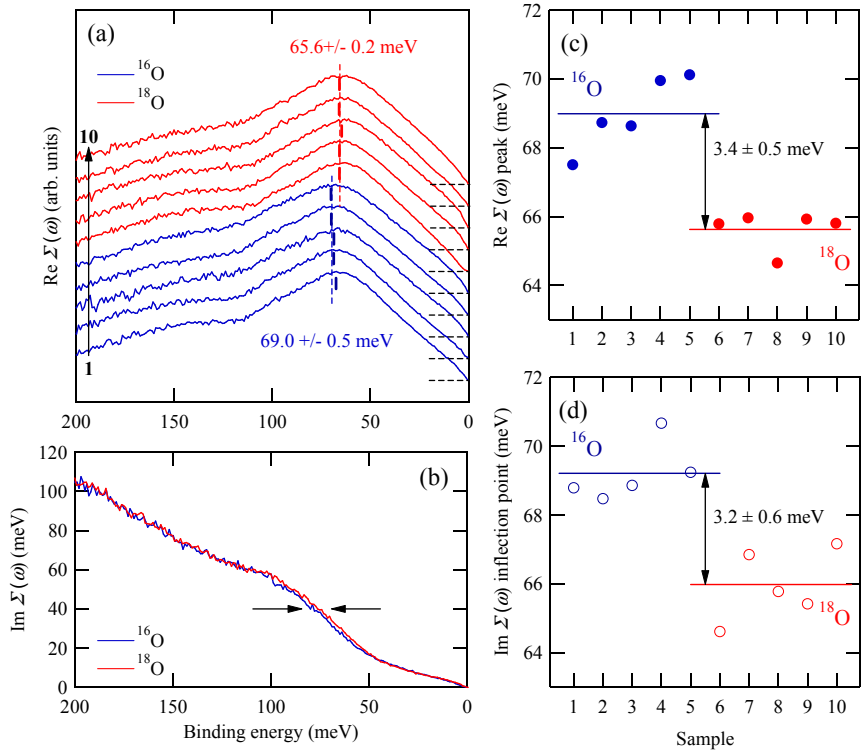}
	\end{center}
	\caption{(color online). (a) Real part of the self-energy Re$\mathit\Sigma$($\mathit\omega$) from five samples both for $^{16}$O (blue) and $^{18}$O (red) along the nodal direction indicated by the red line in Fig.~\ref{FIG1} (c). 
All Re$\mathit\Sigma$($\mathit\omega$) are deduced by subtracting a bare band dispersion from the experimental one, where $\mathit\omega$ is the energy relative to the Fermi energy, and normalized by the peak maximum, and are also offset for clarity. 
(b) Imaginary part of the self energy Im$\mathit\Sigma$($\mathit\omega$) determined from MDC full widths.
An impurity scattering term at $\mathit\omega$ = 0 is subtracted as an energy independent constant background.  
(c) and (d) Obtained kink energy as a function of sample numbers both for $^{16}$O (blue) and $^{18}$O (red) from Re$\mathit\Sigma$($\mathit\omega$) and Im$\mathit\Sigma$($\mathit\omega$), respectively.}	
\label{FIG3}
\end{figure*}

Figure~\ref{FIG3} (a) shows the real part of the self-energy Re$\mathit\Sigma$($\mathit\omega$) both for $^{16}$O (blue) and $^{18}$O (red) multiple samples. 
From these Re$\mathit\Sigma$($\mathit\omega$), we obtained kink energy plotted as a function of five different samples each for both $^{16}$O and $^{18}$O in Fig.~\ref{FIG3} (c).  
We found a clear isotopic softening of the kink energy from about 69.0 meV to about 65.6 meV, or a softening of 3.4 $\pm$ 0.5 meV \cite{IE_com}. 
Additionally, we used a completely independent analysis method using the widths of the ARPES peaks.
This analysis has the advantage of not having any assumptions about a bare band, such that the isotope effect should appear more straightforward.  
Thus we see a 3.2 $\pm$ 0.6 meV shift in the imaginary part of the self-energy Im$\mathit\Sigma$($\mathit\omega$) [Figs.~\ref{FIG3} (b) and (d)]. 
By studying ten samples as well as by using multiple independent analysis methods, we compensated for possible systematic errors which might come into play when trying to determine energies to such a great precision.	
Therefore, we can state with confidence that the $\sim$ 70 meV feature in the nodal electron self-energy is due to the coupling of the electrons with phonons. 
That this is the dominant feature in the electron self-energy, as is seen from both the real [Fig.~\ref{FIG3} (a)] and imaginary [Fig.~\ref{FIG3} (b)] parts of the spectrum, is clear and significant. 

Then, which phonons are responsible for this coupling?
Neutron scattering experiments \cite{Pintschovius06} as well as first-principles phonon calculations \cite{Bohnen03} indicate a few phonon modes that are likely to be most relevant for the coupling; the in-plane $``$half-breathing" phonon mode ($\Omega$ $\sim$ 70 meV) and the $``$buckling/stretching" modes ($\Omega$ $\sim$ 36 meV). 
The apical oxygen stretching mode ($\Omega$ $\sim$  50 meV) could also be considered, though calculations indicate that the number of allowed final states for these phonons is negligible \cite{Giustino08}.
By coupling the nodal electrons with momentum, $\mathbf{k}$, to other parts of the Fermi surface ($\mathbf{k'}$), the electron self-energy can in principle pick up the energy of the superconducting gap $\mathit{\Delta}$($\mathbf{k'}$) weighted over the Brillouin zone (and hence gap size) by the electron-phonon matrix elements $|g$($\mathbf{k}, \mathbf{k'}$)$|$, though there is still theoretical disagreement about how to do this gap-referencing in a $d$-wave superconductor \cite{com4}. 
On the one hand, it is suggested that the kink energy appears at the mode energy plus the maximum gap energy, or $\Omega$+$\mathit{\Delta}_{max}$ \cite{Lee08}. 
This scenario would indicate that the nodal kink is caused by the 36 meV buckling phonon. 
Alternatively, it is suggested that the kink represents simply the mode energy, indicating that the half-breathing phonon is dominant and couples electrons strongly along the node \cite{Giustino08, Heid08, Devereaux04}. 
It is here that the present isotope effect is uniquely capable to distinguish the two. 

For the buckling mode with $\Omega$ $\sim$ 36 meV, the isotope shift would be expected to be 2.1 meV, more than two error bars outside of the results published here. 
On the other hand, the breathing mode with $\Omega$ $\sim$ 69 meV would have an isotope shift of 3.9 meV, within the error bar of the isotopic shift seen here. 
This serves as a strong indication that the breathing mode is responsible for the nodal kink in ARPES data, consistent with expectations from recent theory \cite{Giustino08}. 
However, the moderately strong coupling that we observe, (a coupling parameter $\lambda$ $\sim$ 0.6 \cite{Lambda_com}) is significantly stronger than that calculated by theory \cite{Giustino08, Heid08}, indicating that these calculations are missing an important ingredient to the electron-phonon coupling in the cuprates. 
Nominally we expect such enhancements in the electron-phonon coupling to originate in the strong electron correlation effects \cite{Koikegami08} | a problem which is interesting in its own right, but which also may have particular relevance to the mechanism of superconductivity in the cuprates. 

In summary, we reported the isotopic fingerprint of the nodal kink probed by high-precision LE-ARPES. 
Present isotope shift of the kink energy provides the first convincing and direct evidence that the electron-phonon interactions are responsible for the origin of the nodal kink. 

This work was supported by KAKENHI (19340105), Research Fellowships of the Japan Society for the Promotion of Science for Young Scientists, DOE grant DE-FG02-03ER46066, and Grant-in-Aid for COE research (No. 13CE2002) of MEXT Japan. 
The synchrotron radiation experiments have been done under the approval of HSRC (Proposal No. 06-A-15).


\begin{thebibliography}{99}
\bibitem{Anneal}				We performed annealing under the same thermal conditions in either $^{16}$O$_2$ or $^{18}$O$_2$ gas: 700 $^\circ$C for 168 hours at 1.0 bar, followed by 700 $^\circ$C for 168 hours at 0.2 bar.
\bibitem{Raman}				The isotopic shift in the $B_{1g}$ mode in the CuO$_2$ layer, and also in oxygen vibrational modes in the BiO layer and the SrO layer were observed, giving the estimation of isotope exchange rates.
\bibitem{IE_com}				Isotope shift were found to be insensitive to the choice of the bare velocity; the isotope shift shown here of 3.4 $\pm$ 0.5 meV utilized a linear bare dispersion, while 3.3 $\pm$ 0.5 meV in the case of a second order polynomial.
\bibitem{Lambda_com}		The ratio of the $``$bare" dispersion in the absence of the kink to the renormalized dispersion due to the kink is $v_{bare}$/$v_{Fermi}$ = 1+ $\lambda$ $\sim$ 1.6.
\end{thebibliography}

\end{document}